\documentclass[twocolumn,preprintnumbers,amsmath,revsymb,nofootinbib]{revtex4}

\usepackage{amsfonts}
\usepackage{amssymb}
\usepackage{wasysym}
\usepackage{mathrsfs}
\usepackage{fontenc}
\usepackage{graphicx}
\usepackage{dcolumn}
\usepackage{bm}
\usepackage{feynmf}
\usepackage{verbatim}
\usepackage{hyperref}
\usepackage{subfigure}
\usepackage{enumerate}
\usepackage{slashed}
\usepackage{array}
\usepackage{color}

\newcommand{\ds}{\displaystyle×}
\newcommand{\ket}{\rangle}
\newcommand{\bra}{\langle×}
\newcommand{\tev}{\,{\rm TeV}}
\newcommand{\gev}{\,{\rm GeV}}

\newcommand{\fbinv}{\rm fb$^{-1}$}

\newcommand{\eg}{{e.g.,}\ }
\newcommand{\ie}{{i.e.,}\ }
\newcommand{\beq}{\begin{equation}}
\newcommand{\eeq}{\end{equation}}

\newcommand{\eqn}{{Eq.}\ }
\newcommand{\trm}[1]{\textrm{#1}}

\newcommand{\thickhline}{%
    \noalign {\ifnum 0=`}\fi \hrule height 1pt
    \futurelet \reserved@a \@xhline
}

\begin{document}

\preprint{ACFI-T14-14}

\title{Singlet-Catalyzed Electroweak Phase Transitions and Precision Higgs Studies}
\author{Stefano Profumo$^{1}$, Michael J. Ramsey-Musolf$^{2,3}$, Carroll L. Wainwright$^{1}$, Peter Winslow$^{2}$}
\affiliation{$^{1}$Santa Cruz Institute for Particle Physics, University of California, Santa Cruz, CA 95064, USA}
\affiliation{$^{2}$Physics Department, University of Massachusetts Amherst, Amherst, MA 01003, USA}
\affiliation{$^{3}$Kellogg Radiation Laboratory, California Institute of Technology, Pasadena, CA 91125, USA}

\date{\today}

\begin{abstract}
We update the phenomenology of gauge singlet extensions of the Standard Model scalar sector and their implications for the electroweak phase transition. Considering the introduction of one real scalar singlet to the scalar potential,  we analyze present constraints on the potential parameters from Higgs coupling measurements at the Large Hadron Collider (LHC) and electroweak precision observables for the kinematic regime in which no new scalar decay modes arise. We then show how future precision measurements of Higgs boson signal strengths and Higgs self-coupling could probe the scalar potential parameter space associated with a strong first-order electroweak phase transition. We illustrate using benchmark precision for several future collider options, including the High Luminosity LHC (HL-LHC), the International Linear Collider (ILC), TLEP, China Electron Positron Collider (CEPC), and a 100 TeV proton-proton collider, such as the Very High Energy LHC (VHE-LHC) or the Super proton-proton Collider (SPPC). For the regions of parameter space leading to a strong first order electroweak phase transition, we find that there exists considerable potential for observable deviations from purely Standard Model Higgs properties at these prospective future colliders.

\end{abstract}

\maketitle

\section{Introduction}\label{Intro}



The ultimate structure of electroweak symmetry breaking (EWSB) is a key question in elementary particle physics. Consequentially, after the recent discovery of a Higgs particle at the Large Hadron Collider (LHC)~\cite{:2012gu,:2012gk}, detailed measurements of its properties have become one of the main priorities of the LHC and future colliders. Thus far, measurements of its couplings are consistent with those expected from the Standard Model (SM). Moreover, including the mass of the new particle in the global electroweak SM fit to precision observables now asserts the validity of the SM at the 0.1\% level~\cite{Baak:2012kk}. Despite this, the SM cannot yet be complete. One outstanding problem, which lies at the interface of cosmology with particle and nuclear physics, is to explain the origin of the baryon asymmetry of the universe:
\begin{eqnarray}
\frac{n_B}{s} = (8.59 \pm 0.11) \times 10^{-11} \quad \trm{(Planck)~\cite{Ade:2013zuv}}
\end{eqnarray}
where $n_B$ ($s$) is the baryon number (entropy) density. Among the several proposed scenarios, electroweak baryogenesis (EWBG) is particularly interesting as it is inextricably tied to electroweak symmetry breaking (for a recent review see, \eg\cite{Morrissey:2012db}) and therefore potentially subject to current and near-future collider probes of the scalar sector. At the same time, current and future searches for the permanent electric dipole moments of atoms, molecules, and nucleons provide a powerful window on the CP-violation needed for baryon asymmetry generation.

Successful EWBG requires a strongly first-order electroweak phase transition (EWPT) that proceeds via bubble nucleation at temperatures $\mathcal{O}$(100 \gev) and new sources of CP violation. CP-violating interactions at the bubble walls induce chiral asymmetries which bias rapid sphaleron processes in the unbroken phase to create non-zero baryon density. This baryon density subsequently diffuses into the broken phase where EWSB suppresses the rate of sphaleron transitions, freezing in a baryon asymmetry. If not sufficiently suppressed, the sphalerons will restore equilibrium by washing out the asymmetry altogether. The strength of the first-order EWPT necessary to avoid washout is generally characterized as $\phi(T_c) / T_c \gtrsim 1$, where $\phi$ is the $SU_L(2)$ scalar background field and $T_c$ is the critical temperature at which the free energies of the broken and unbroken phases become degenerate\footnote{As we discuss below, one must exercise care in defining this ratio in order to maintain gauge invariance.}. In the SM, this condition can only be satisfied for values of the Higgs mass far below the value indicated by the LHC.
However, in extensions of the SM, with more complex scalar sectors, this condition for a strong first order EWPT (SFOEWPT) can be readily met. 

In this work, we revisit the viability of a SFOEWPT in the simplest extension of the SM scalar sector involving one real gauge singlet scalar $S$, dubbed the ``xSM'' , and analyze its implications for future, precision Higgs studies (for earlier studies of the EWPT dynamics and/or phenomenology of the xSM, see \eg \cite{O'Connell:2006wi,Barger:2007im,Profumo:2007wc,Noble:2007kk,He:2009yd,Petraki:2007gq,Gonderinger:2009jp,Cline:2009sn,Espinosa:1993bs,Cline:2012hg,Cline:2013gha,Katz:2014bha,No:2013wsa,Fuyuto:2014yia} and references therein). 
This scenario does not contain new sources of CP violation but, in its most general form, provides a framework for studying the generic characteristics of the EWPT in models with extra singlet scalars. It also falls within the class of Higgs portal scenarios, in which the dominant connection between the SM and new physics sectors is through operators of the form $H^\dagger H S$ and $H^\dagger H S^2$. After EWSB, these operators induce mixing between $S$ and $H$, giving rise to a pair of neutral mass eigenstates, $h_{1,2}$. Depending on their masses, the Higgs portal operators can enable either new decay~\cite{O'Connell:2006wi,Barger:2007im,Profumo:2007wc,He:2009yd,Petraki:2007gq,Gonderinger:2009jp} or resonant di-Higgs production~\cite{No:2013wsa} modes. 

In our analysis, we consider the kinematic regime in which no new on-shell decay modes arise and di-Higgs production is non-resonant. As a result, in order to probe the xSM scenario in this regime, precision measurements are required. First, mixing generates small deviations to Higgs production rates. Although current LHC measurements constrain deviations to roughly $\mathcal{O}$(20\%)~\cite{ATLASupdate}, future collider experiments such as the high-luminosity LHC (HL-LHC), the International Linear Collider (ILC), TLEP, China Electron Positron Collider (CEPC), and a 100 TeV proton-proton collider, such as the Very High Energy LHC (VHE-LHC) or Super Proton Proton Collider (SPPC) are expected to significantly improve the accuracy of these measurements. Second, mixing can lead to significant deviations of the Higgs trilinear self-coupling from its SM value, leading to potentially observable consequences for self-coupling studies. Finally, the presence of another scalar state can be probed indirectly, by precision electroweak observables, and directly, by searches for singlet-like heavy Higgs bosons in the low mass region, $<2 m_h$. We consider all these constraints as well as the projected sensitivity of future collider experiments.

As Higgs portal interactions can also enable the occurrence of a SFOEWPT~\cite{Profumo:2007wc}, the aforementioned experimental signatures provide a link between collider phenomenology and phase transition dynamics. In our analysis, we rely on a combination of analytic and numerical methods to analyze the xSM EWPT and its phenomenological consequences. We first revisit analytic calculations of the strength, relying heavily on Ref.~\cite{Profumo:2007wc}, in order to gain intuition of the generic characteristics. Then, up-dating that work, we rely on the {\tt CosmoTransitions} package~\cite{Wainwright:2011kj} to calculate various aspects of the EWPT numerically. Our strategy in this work is to ascertain where current and future collider searches would probe the SFOEWPT-viable parameter space. 

Our analysis indicates that a SFOEWPT prefers large negative couplings associated with the $H^\dagger H S$ Higgs portal operator. As we discuss below, this preference biases the associated collider phenomenology towards large mass splittings between the scalar eigenstates in the region of small mixing angles while allowing for potentially significant reductions in the strength of the SM-like scalar self coupling. Future precision measurements of the SM-like scalar signal strengths and self-coupling would, then, provide powerful probes of the SFOEWPT-viable parameter space. In this respect, there appears to be considerable potential for observable deviations from purely SM-like Higgs properties. Moreover, direct searches for singlet-like scalars having SM-like Higgs branching ratios but reduced signal strengths would provide an additional window on this scenario. Combining such searches with precision Higgs property measurements at future colliders could, thus, reveal the presence of scalar potential dynamics needed for preserving any baryon asymmetry produced during the EWSB era\footnote{We note that, owing to our implementation of gauge independence in the {\tt CosmoTransitions} package, we do not consider the region of parameter space where a first order EWPT may arise through the combination of loop-induced SM contributions and an effective reduction in the Higgs quartic self-coupling as observed in Ref.~\cite{Profumo:2007wc}.}.

Our discussion of this analysis is organized as follows: in Section II, we establish our notations for the xSM model and discuss basic theoretical bounds. Section III describes our fit to the current Higgs coupling measurements and study of future sensitivities of the HL-LHC, ILC, TLEP, CEPC, and VHE-LHC or SPPC. In this section, we also present constraints from heavy SM-like Higgs searches and electroweak precision observables. In Section IV, we discuss the finite temperature effective potential and our analysis of the EWPT. In section V, we present our final comments and conclusions. 


\section{The xSM: A Singlet Scalar Extension of the SM}\label{sec:Model}


We study a minimal extension of the SM scalar sector consisting of a single, gauge singlet, real scalar field S. The $T=0$ tree level potential for the Higgs doublet $H$ and $S$ is given by 
\begin{eqnarray}
& V_0^{T=0}(H,S) =\ds  -\mu^2 \left( H^\dagger H \right) + \lambda \left( H^\dagger H \right)^2 + \frac{a_1}{2} \left( H^\dagger H \right) S & \nonumber \\
& \ds + \frac{a_2}{2} \left( H^\dagger H \right) S^2 + \frac{b_2}{2} S^2 + \frac{b_3}{3} S^3 + \frac{b_4}{4} S^4 . &
\label{TotPot}
\end{eqnarray}
The $a_1$ and $a_2$ parameters constitute the Higgs portal which provides the only connection to the SM for the singlet scalar $S$. The $b_2$, $b_3$, and $b_4$ parameters are self-interactions which, without the Higgs portal, constitute a hidden sector. Our notation here follows that of Refs.~\cite{O'Connell:2006wi,Profumo:2007wc,Barger:2007im}, where no distinction is made between dimensionful and dimensionless couplings. Modulo the $a_1$ and $b_3$ parameters, the potential has a $\mathbb{Z}_2$ symmetry that stabilizes the singlet scalar, enabling a dark matter interpretation\footnote{As was pointed out in Ref.~\cite{Profumo:2007wc}, even if a $\mathbb{Z}_2$ symmetry is present before electroweak symmetry breaking, both scalar eigenstates are made unstable through mixing if $\bra S \ket \neq 0$.} which has been studied by many previous authors, \eg \cite{Gonderinger:2009jp,Cline:2012hg,Cline:2013gha}. However, as these parameters play a large role in the strength of the EWPT, we retain them, thereby rendering $S$ incapable of simultaneously providing a successful dark matter candidate. 

As we are interested in the general pattern of EWSB for non-vanishing temperatures, we let $S$, as well as $H$, take on a vacuum expectation value (vev), \ie $S \to x_0 + s$ and $H \to (v_0+h) / \sqrt{2}$ where $x_0$ and $v_0$ are the $T=0$ vevs. Since $x_0$ also breaks $\mathbb{Z}_2$ symmetry, we choose it to be positive through a field redefinition ($s \to -s$). The minimization conditions then allow us to express two of the potential parameters in \eqn (\ref{TotPot}) in terms of the $T=0$ vevs and other potential parameters as
\begin{eqnarray}
& \ds \mu^2 = \lambda v_0^2 + \left( a_1 + a_2 x_0 \right) \frac{x_0}{2} & \nonumber \\
\label{eq:ewsb}
& \ds b_2 = \ds - b_3 x_0 - b_4 x_0^2 - \frac{a_1 v_0^2}{4 x_0} - \frac{a_2 v_0^2}{2} .&
\end{eqnarray}
We find it useful to exchange these two mass dimension two parameters for those appearing on the RHS of Eq.~(\ref{eq:ewsb}) that have mass dimension one or zero. Doing so is particularly advantageous for numerical scans as we may choose smaller ranges for the latter parameters than would otherwise be necessary for the mass-squared parameters.

The elements of the tree level mass-squared matrix are given by
\begin{eqnarray}
& m_{hh}^2 \equiv \ds \frac{d^2 V}{dh^2} = 2 \lambda v_0^2 & \nonumber \\
& m_{ss}^2 \equiv \ds \frac{d^2 V}{ds^2}  = b_3 x_0 + 2 b_4 x_0^2 - \frac{a_1 v_0^2}{4 x_0} & \nonumber \\
& m_{hs}^2 \equiv \ds \frac{d^2 V}{dh ds} = \left(a_1 + 2 a_2 x_0 \right) \frac{v_0}{2} ,
\label{mixingM}
\end{eqnarray}
with the corresponding mass eigenstates
\begin{eqnarray}
& h_1 = h \cos \theta + s \sin \theta & \nonumber \\
& h_2 = - h \sin \theta + s \cos \theta &
\label{eigstates}
\end{eqnarray}
where $h_1$ ($h_2$) is the more $SU_L(2)$-like (singlet-like) scalar. The mixing angle $\theta$ is most easily defined in terms of the mass eigenvalues,
\begin{eqnarray}
& m_{1,2}^2 = \ds \frac{ m_{hh}^2 + m_{ss}^2 \pm \left| m_{hh}^2 - m_{ss}^2 \right| \sqrt{ 1 + \ds \left( \frac{ m_{hs}^2 }{ m_{hh}^2 - m_{ss}^2 } \right)^2 } } {2} ,& \nonumber \\
\label{Meigenvalues}
\end{eqnarray}
as
\begin{eqnarray}
\sin 2 \theta =  \ds \frac{ \left( a_1 + 2 a_2 x_0 \right) v_0  }{ \left( m_1^2 - m_2^2 \right) } .
\label{sin2theta}
\end{eqnarray}
Here, we make several points.
\begin{itemize}
\item[$\bullet$] We require that the $SU_L(2)$-like scalar eigenstate, $h_1$, is the lighter eigenstate and identify it with the observed Higgs boson at the LHC~\cite{:2012gu,:2012gk}, \ie we set $m_1 \equiv 125 \gev$. The couplings to all SM states are then rescaled by $\cos\theta$, which results in modifications to the production cross sections that can be constrained by measurements of Higgs signal strengths. We explore these constraints in the next section.
\item[$\bullet$] The singlet-like scalar eigenstate, $h_2$, receives its decay modes entirely from mixing, via the Higgs portal couplings $a_1$ and $a_2$. As such, these parameters are also sensitive to collider searches for heavy SM-like Higgs bosons. Moreover, electroweak precision observables are sensitive to the presence of new heavy scalar states. We explore the effect of these constraints on the mixing angle, $\theta$, and mass, $m_2$, in the next section.
\item[$\bullet$] The relation for the mixing angle $\theta$ in \eqn (\ref{sin2theta}) implies a highly non-trivial bound on the Higgs portal parameters and physical masses,
\begin{eqnarray}
-1 \leq \ds \frac{ \left( a_1 + 2 a_2 x_0 \right) v_0  }{ \left( m_1^2 - m_2^2 \right) } \leq 1, 
\label{thetabound}
\end{eqnarray}
which becomes more severe in the limit in which $h_1$ and $h_2$ are degenerate.
\end{itemize}
%
%
To avoid vacuum instability at $T=0$, the potential in \eqn~(\ref{TotPot}) must be bounded from below. This is imposed by requiring the positivity of the quartic coefficients along all directions in field space. Along the $h$ ($s$) direction, this leads to the bound $\lambda > 0$ ($b_4 > 0$) while, along an arbitrary direction, this implies $a_2 > - \sqrt{\lambda b_4}$. 

For viable EWSB, two conditions must be met. The first is that the determinant of the mass mixing matrix described in \eqn (\ref{mixingM}) must be positive for $T \leq T_c$, where $T_c$ is the critical temperature associated with the EWPT. This is true when
\begin{eqnarray}
\ds b_3 x + 2 b_4 x^2 - \frac{a_1 v^2}{4 x} - \frac{ (a_1 + 2 a_2 x )^2 }{ 8 \lambda } > 0 ,
\end{eqnarray}
where $v$ and $x$ denote the finite temperature vevs for $T \leq T_c$. The second condition is that the electroweak minimum must be the absolute minimum at $T=0$, which we impose numerically.\\

\section{Collider Phenomenology and Electroweak Precision Observables}\label{sec:ColliderPhenoEW}


Phenomenologically, current measurements of Higgs couplings constrain the combinations of potential parameters that determine the singlet-like scalar mass eigenvalue, $m_2$, and the mixing angle $\sin 2 \theta$. In this work, we concentrate on the kinematic regime in which no new scalar decay modes arise, \ie $m_1 /2 < m_2 \leq 2 m_1$, where we remind the reader that we have defined $m_1 = 125 \gev$. This scenario is particularly challenging experimentally and the strategy to probe it necessarily must include high precision measurements of Higgs couplings. Motivated by this, we study not only the current status of LHC measurements of Higgs couplings but also projections for experiments at the HL-LHC, ILC, TLEP, CPEC, and VHE-LHC or SPPC as well. Moreover, we include in our analysis SM-like Higgs searches in the low mass regime and electroweak precision observables. \\

\begin{figure*}[t!]
\centering
\hspace*{-.1in}\mbox{
\subfigure{
\includegraphics[scale=.615]{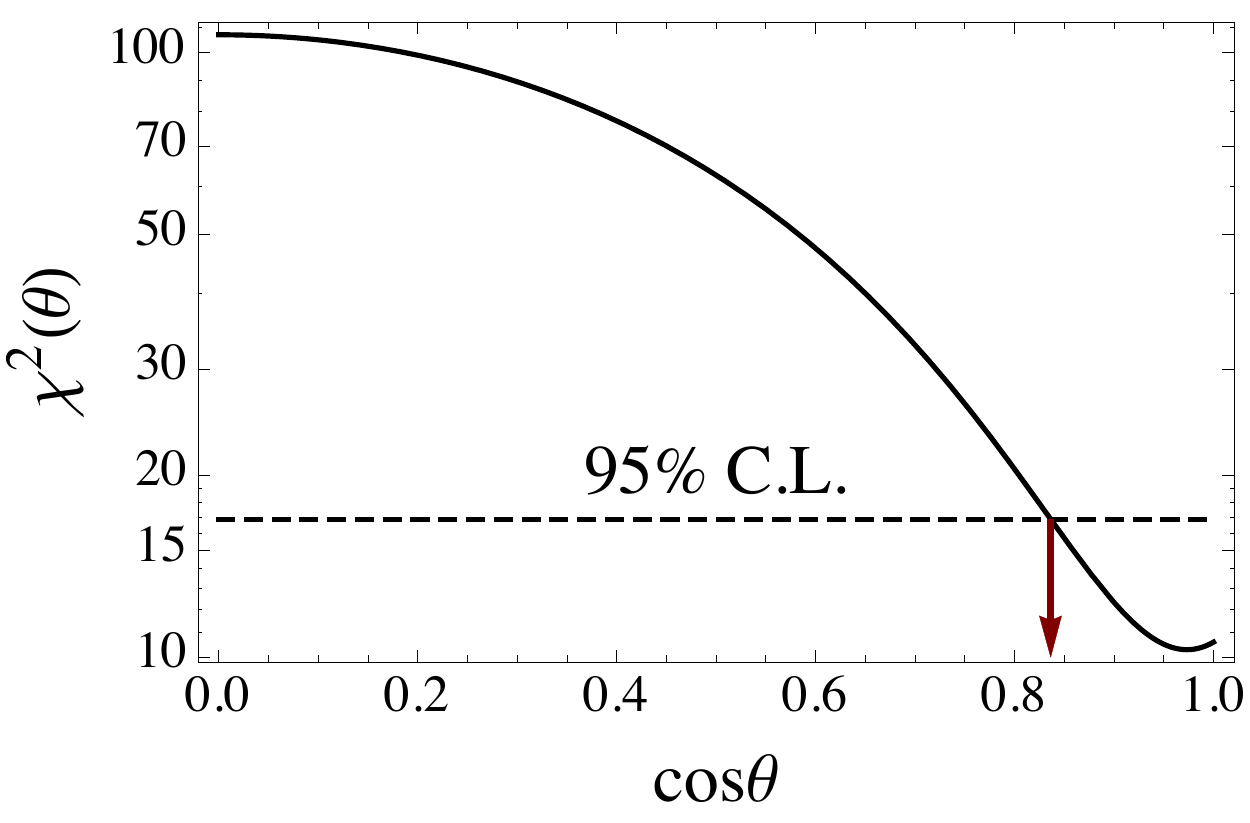}
}
\quad \quad \quad \subfigure{
\includegraphics[scale=.625]{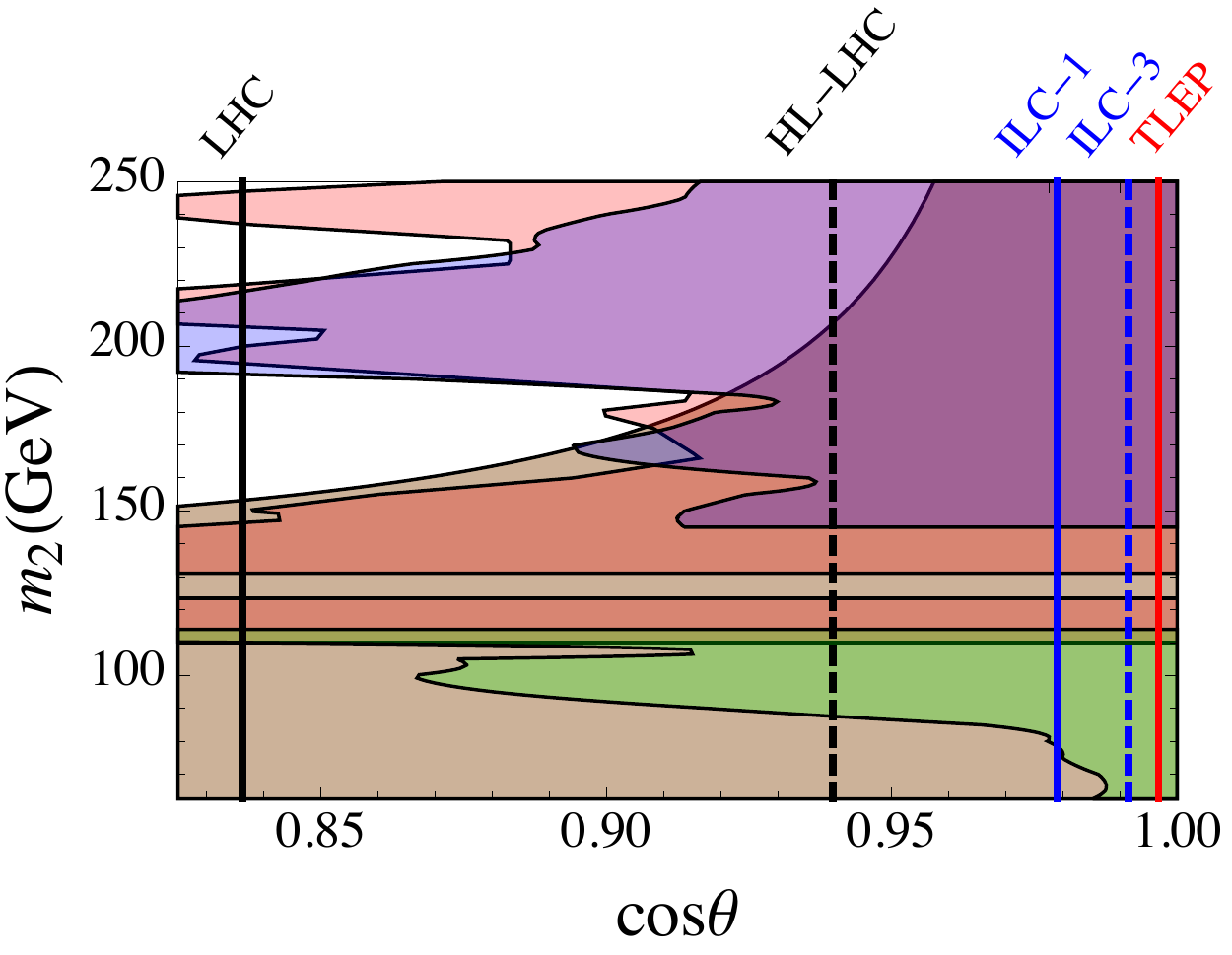}
}
}
\caption{Left panel: \textit{$\chi^2$ fit to mixing angle from current Higgs measurements at the LHC. The current 95\% C.L. limit is $|\cos \theta| \geq 0.84$.} Right panel: \textit{ The combined current and projected constraints on the mixing angle and singlet-like eigenstate mass $m_2$. The current limits are from a combined fit to current ATLAS and CMS Higgs measurements (black solid line), CMS heavy SM-like Higgs searches (blue region), null results from LHC SM Higgs searches (red region), null results from LEP SM Higgs searches (green region), and electroweak precision observables (beige region). The projected constraints on the mixing angle are from the $\sqrt{s}=14 \tev$, $3 \; ab^{-1}$ HL-LHC run (black dashed line), $\sqrt{s}=250 \gev$, $250 \; fb^{-1}$ (ILC-1) ILC run (blue solid line), $\sqrt{s}=1 \tev$, $1\; ab^{-1}$ (ILC-3) ILC run (blue dashed line), and $\sqrt{s}=240 \gev$, $1\; ab^{-1}$ TLEP run (red solid line).}}
\label{FIG2}
\end{figure*}

From \eqn (\ref{eigstates}), the couplings of the $SU_L(2)$-like eigenstate, $h_1$, to all SM states are simply rescaled versions of SM Higgs couplings,
\begin{eqnarray}
g_{h_1 XX} = \cos \theta g_{hXX}^{SM} .
\label{h1Coupling}
\end{eqnarray}
Since $m_1$ is fixed, all signal rates $\mu_{XX}$ associated with Higgs measurements, relative to pure SM-Higgs expectations, are strictly functions of the mixing angle as 
\begin{eqnarray}
\mu_{XX} = \frac{ \sigma \cdot \trm{BR} }{ \sigma^{SM} \cdot \trm{BR}^{SM} } = \cos^2 \theta ,
\end{eqnarray}
where $\sigma$ is the production cross section and the branching ratios, $\trm{BR}$, are independent of the mixing angle as there are no new additions to the total width. We then impose constraints on the mixing angle by performing a global $\chi^2$ fit to the current Higgs data from both ATLAS~\cite{ATLASupdate} and CMS~\cite{Chatrchyan:2013zna,Chatrchyan:2014nva,CMS:ril,Chatrchyan:2013iaa,Chatrchyan:2013mxa} using 
\begin{eqnarray}
\chi^2 ( \theta ) = \sum_i  \left( \frac{ \mu_i^{obs} - \cos^2 \theta }{ \Delta \mu_i^{obs}  } \right)^2 ,
\end{eqnarray}
where $\mu_i^{obs}$ ($\Delta \mu_i^{obs}$) are the (uncertainties in the) observed signal rates. The result, shown in the left panel of Fig. \ref{FIG2}, is that the present 95\% C.L. limit on the mixing angle is $|\cos \theta| \geq 0.84$.
 
In principle, it is possible to obtain estimates of sensitivities to the mixing angle from future collider experiments. Loosely following Ref.~\cite{Peskin:2013xra}, we adopt a ``naive $\chi^2$ method'' in which we take as input the projected estimates for uncertainties of signal rates, $\Delta \mu^{proj}$, presented by experimental collaborations. We then assume each measurement to be independent, centered on the SM expectation, and gaussian distributed so that we may extract fit values of $\cos \theta$ from the $\chi^2$ function
\begin{eqnarray}
\chi^2 ( \theta ) = \sum_i  \left( \frac{ 1 - \cos^2 \theta }{ \Delta \mu_i^{proj}  } \right)^2 .
\end{eqnarray}
For the HL-LHC, the ATLAS collaboration has provided such projected uncertainties on its signal rate measurements at $\sqrt{s}=$14 \tev~for both 300 \fbinv~and 3000 \fbinv~\cite{ATLASproj}, assuming current theoretical and systematic uncertainties. The CMS collaboration has also presented a set of estimates at $\sqrt{s}=$14~\tev~for both 300 \fbinv~and 3000 \fbinv, using two scenarios~\cite{CMS:2013xfa}. In scenario 1, current systematic and theoretical uncertainties are left unchanged. In scenario 2, theoretical uncertainties are taken to be $1/2$ their current value while systematic uncertainties are scaled by the square root of the luminosity. In an effort to take values whose underlying assumptions match those of the ATLAS analysis as much as possible, we take inputs from scenario 1 only. Moreover, we find that the projected limit on $\theta$ does not vary much between 300 \fbinv~and 3000 \fbinv, so we simply show limits for the most ambitious scenario of 3000 \fbinv. We then interpret the projected HL-LHC sensitivity to the mixing angle as the 95\% C.L. limit from the combined fit to both ATLAS and CMS projections. The result is presented in the right panel of Fig. \ref{FIG2} as a black dashed vertical line.

While precision measurements at the HL-LHC are made difficult by rising systematic uncertainties, ILC uncertainties will be dominated by statistical errors and so are expected to continually improve with more data. In the ILC Higgs white paper~\cite{Asner:2013psa}, estimates for uncertainties in signal rates are given at 3 separate stages, $\sqrt{s}=$250 \gev~for 250 \fbinv (ILC-1), $\sqrt{s}=$500 \gev~for 500 \fbinv (ILC-2), and $\sqrt{s}=$1000 \gev~for 1000 \fbinv (ILC-3). Using the same naive $\chi^2$ method, we produce estimates for the sensitivity to the mixing angle from the various stages of the ILC collider program. Again, as the variation in sensitivity between the ILC-1 and ILC-2 programs or ILC-2 and ILC-3 programs is quite small, we only present the projected sensitivities from the ILC-1 and ILC-3 programs. The results are presented in the right panel of Fig. \ref{FIG2}, as solid (ILC-1) and dashed (ILC-3) blue vertical lines.

The recently proposed TLEP collider is an $e^+ e^-$ circular collider capable of high luminosities with a center of mass energy range between 90 and 500 \gev~\cite{Koratzinos:2013ncw}. Estimated uncertainties for signal rates have also been quantified for TLEP assuming $\sqrt{s}=$240 \gev, optimal for Higgs production, and 1 ab$^{-1}$~\cite{Gomez-Ceballos:2013zzn}. Employing the same naive $\chi^2$ method, we calculate the sensitivity to the mixing angle and present the result in the right panel of Fig. \ref{FIG2} as a solid red line. The sensitivities for the CEPC are expected to be similar to those for TLEP, so we will use the TLEP benchmarks in Ref.~\cite{Koratzinos:2013ncw} as indicative for both collider options.

We also note here that such estimates of signal rate uncertainties are, to the best of our knowledge, not yet available for the VHE-LHC or SPPC, so we cannot yet estimate its sensitivity to the mixing angle.

Aside from precision studies of Higgs signal rates, another way to experimentally probe the xSM scenario is by directly searching for the heavy singlet-like mass eigenstate, $h_2$. This state inherits all its interactions with the SM entirely from the mixing and so, modulo a rescaling of the couplings
\begin{eqnarray}
g_{h_2 XX} = \sin \theta g_{hXX}^{SM} ,
\label{h2Coupling}
\end{eqnarray}
has a similar phenomenology to that of the SM Higgs. In particular, our expectations of branching ratios should be unchanged. In direct analogy with the previous analysis, the signal rates relative to the SM rates are independent of the $h_2$ mass and functions of the mixing angle only, \ie
\begin{eqnarray}
\mu_{XX} = \frac{ \sigma (m_2) \cdot \trm{BR} (m_2) }{ \sigma^{SM} (m_2) \cdot \trm{BR}^{SM} (m_2) } = 1- \cos^2 \theta .
\end{eqnarray}
As the mass of $h_2$ approaches the di-boson thresholds, $2 M_W$ and $2 M_Z$, the branching ratios to these states begin to dominate. Motivated by this, the CMS collaboration has performed a search for a SM-like Higgs in the mass range $145-1000$ \gev, concentrating on the $WW$ and $ZZ$ final states~\cite{Chatrchyan:2013yoa}. They place a limit on a signal rate for the heavy SM-like boson by normalizing the observed rate to the SM prediction for the rate as a function of the Higgs mass. The corresponding constraint is presented in the right panel of Fig. \ref{FIG2}, with the blue region representing the allowed region. 

Null results from SM Higgs searches~\cite{:2012gk} also probe the presence of a second scalar state. Although the mass range for these bounds extend from 110 GeV to 600 GeV, we study only the range up to $2 m_h$. We present the allowed region from these bounds as the shaded red region in the right panel of Fig. \ref{FIG2}. The mass gap at $\sim$125 GeV cannot be excluded due to the observed presence of the $SU_L(2)$-like scalar eigenstate. Moreover, once $m_2 <$114 GeV, the presence of such a low mass scalar is subject to tight bounds from LEP~\cite{Barate:2003sz}. These bounds constrain the mixing angle through null searches for Higgs-strahlung  production of the singlet-like scalar, $h_2$. We present the allowed region from LEP bounds in the right panel of Fig. \ref{FIG2} as the green shaded region. We emphasize  that the signal rates relative to the SM rates have no mass dependence. However, the limits themselves are mass-dependent, and it is this mass-dependence that is seen in the right panel of Fig. \ref{FIG2}.

Finally, we consider the impact of electroweak precision observables on the xSM by computing the scalar contributions to the diagonal, weak gauge boson vacuum polarization diagrams\footnote{As real neutral scalars carry no electric charge, they have no effect on the $\Pi_{\gamma \gamma}$ and $\Pi_{\gamma Z}$ polarization amplitudes.}. Using these results, we characterize the effects of the xSM on electroweak precision observables and $W$-boson mass with the oblique parameters $S$, $T$, and $U$. Both $h_1$ and $h_2$ interact with the $W$ and $Z$ via rescaled versions of the corresponding SM Higgs couplings, see \eqn (\ref{h1Coupling}) and (\ref{h2Coupling}). As such, we may write the shift in any oblique parameter, $\mathcal{O}$, entirely in terms of the SM Higgs contribution to that parameter, $\mathcal{O}^{SM}(m)$, where $m$ represents either $m_1$ or $m_2$. The shifts in the oblique parameters, $\Delta \mathcal{O} \equiv \mathcal{O} - \mathcal{O}^{SM}$, then take on the simple form
\begin{eqnarray}
& \Delta \mathcal{O} = \cos^2 \theta \mathcal{O}^{SM}(m_1) + \sin^2 \theta \mathcal{O}^{SM}(m_2) - \mathcal{O}^{SM}(m_1) & \nonumber \\
&  = \left(1 - \cos^2 \theta \right) \left( \mathcal{O}^{SM}(m_2) - \mathcal{O}^{SM}(m_1) \right) ,&
\end{eqnarray}
which makes it clear that the corresponding constraints are significantly weakened for small singlet-like masses, \ie $m_2 \sim m_1$, and small mixing angles. 

In order to implement the fit to electroweak precision observables, we take the best fit values and standard errors for the shifts, $\Delta \mathcal{O}$, from the most recent post-Higgs-discovery electroweak fit to the SM performed by the Gfitter group~\cite{Baak:2012kk}. The following results were obtained for the SM reference point with the top quark (Higgs) mass of $m_{t,ref} = 173 \gev$ ($m_{h,ref} = 126 \gev$), yielding
\begin{eqnarray}
& S - S_{SM} = 0.03 \pm 0.10 & \nonumber \\
& T - T_{SM} = 0.05 \pm 0.12 & \nonumber \\
& U - U_{SM} = 0.03 \pm 0.10 .&
\label{obliques}
\end{eqnarray}
We then define a $\Delta \chi^2$ as 
\begin{eqnarray}
\Delta \chi^2 = \sum_{i,j} \left( \Delta \mathcal{O}_i - \Delta \mathcal{O}_i^0 \right) \left( \sigma^2 \right)^{-1}_{ij} \left( \Delta \mathcal{O}_j - \Delta \mathcal{O}_j^0 \right),
\end{eqnarray}
where $\Delta \mathcal{O}^0_i$ denotes the central values for the shifts in \eqn (\ref{obliques}); $\sigma^2_{ij} \equiv \sigma_i \rho_{ij} \sigma_j$ with $\sigma_i$ denoting the errors in \eqn (\ref{obliques}); and the correlation matrix is~\cite{Baak:2012kk} 
\begin{eqnarray}
\rho_{ij} = \left(
\begin{array}{ccc}
1 & 0.891 & -0.540 \\
0.891 & 1 & -0.803 \\
-0.540 & -0.803 & 1
\end{array}
\right) .
\end{eqnarray}
Following Ref.~\cite{PDG2012}, we take the 95\% C.L. ellipsoid in the space of $\Delta \mathcal{O}_i$ to correspond to $\Delta \chi^2 \leq 5.99$ and interpret parameter regions as consistent with electroweak precision observables if they lie within this region. We present the allowed region of the $\cos \theta$, $m_2$ parameter space in the right panel of Fig. \ref{FIG2} as the beige shaded region. \\



\section{The Electroweak Phase Transition}\label{sec:EWPT}


The standard analysis of the finite temperature effective potential, $V_{eff}^{T \neq 0}$, entails the addition of three separate contributions to the tree level $T=0$ potential: the $T=0$ Coleman-Weinberg 1-loop effective potential; the $T \neq 0$ 1-loop corrections; and the bosonic ring corrections that re-sum  contributions from non-zero Matsubara modes via inclusion of thermal masses in the 1-loop propagators~\cite{Gross:1980br,Parwani:1991gq} (see, \eg Ref.~\cite{Quiros:1999jp}, for a pedagogical review). However, as discussed in depth in Ref.~\cite{Patel:2011th}, the resulting $V_{eff}^{T \neq 0}$ is gauge-dependent. The value of the EWSB vev is inherently gauge-dependent as it is not an observable, while the standard procedure for extracting the critical temperature, $T_c$, introduces a spurious gauge-dependence in the computed value of $T_c$. Thus, the conventional criterion for avoiding baryon washout, $\phi(T_c)/T_c\gtrsim 1$ as na\"ively applied inherits both sources of gauge-dependence. 

There exist various strategies for defining a gauge-invariant baryon number preservation criterion (BNPC). For scenarios where there exist tree-level extremal configurations, one may consistently implement loop corrections in a gauge-invariant manner by employing an $\hbar$-expansion. This approach is particularly applicable when the existence of a first-order EWPT arises from a loop-induced barrier between the broken and unbroken extrema. For the present case, the cubic Higgs portal operator generates a tree-level barrier, suggesting a simpler strategy that we adopt here: 
we forgo the addition of the $T=0$ Coleman-Weinberg 1-loop effective potential and retain only the gauge-independent thermal mass corrections to the effective potential. The latter are decisive for the restoration of EW symmetry at high-temperature, as they eventually overcome the negative mass-squared terms in the potential\footnote{By themselves, the thermal mass corrections induce a second order EWSB transition; the presence of the barrier arising from the tree-level cubic portal operator makes the transition first order.}. For a discussion of the limits of validity of this high-$T$ effective theory, see Ref.~\cite{Carson:1989rf}  and references therein.

This approach allows us to derive a gauge-independent $T \neq 0$ effective potential from which we can extract physically meaningful results. In particular, both the $T$-dependent vevs as well as the computed critical temperature are manifestly gauge-independent. A similar statement applies to the bubble nucleation rate (see below). We note, however, that our approach will not allow us to investigate regions of parameter space where the loop-induced cubic term in the potential drives the first-order EWPT and where suitable choices of the other model parameters ensures that it is suitably \lq\lq strong".

We also note that, with the level of complication involved in the xSM potential, it is possible that the transition to the electroweak phase could have proceeded through multiple steps, passing through intermediary phases before reaching the electroweak phase. The study of Ref.~\cite{Profumo:2007wc}, for example, found that more effective baryon number preservation occurs when the EWPT begins from a phase in which the singlet field has a non-zero vev. Multistep transitions have also been explored in scalar sector extensions involving non-trivial SU(2) scalar representations (\eg see Ref.~\cite{Patel:2012pi}). In principle they may be exploited to generate the baryon asymmetry during a higher-temperature, less experimentally constrained transition that proceeds prior to the transition to the present vacuum. In our current analysis, we will not concern ourselves with the details of the full thermal history of the potential but, instead, remain focused on the implications of the xSM for the EWPT only.

With these caveats in mind, we now study the implications of the xSM model for the strength of the EWPT in the early universe. We begin by following Refs.~\cite{Profumo:2007wc,Pietroni:1992in} and work with a cylindrical coordinate representation for the $T$-dependent vevs
\begin{eqnarray}
& {\bar v}(T)/\sqrt{2} = {\bar \phi}(T) \cos \alpha(T) & \nonumber \\
& {\bar x}(T) ={\bar \phi}(T) \sin \alpha(T) \ \ \ ,&
\end{eqnarray}
where the bar over $v$ and $x$ indicate the are the gauge-independent doublet and singlet vevs in the high-$T$ effective theory. The energy of the electroweak sphaleron responsible for baryon washout is proportional to the SU(2$)_L$-breaking energy scale that, in our high-$T$ effective theory, is given by ${\bar v}(T)$. 
To ensure that electroweak sphalerons are sufficiently quenched in the broken electroweak phase to prevent the washout of any baryon asymmetry, one finds the approximate requirement: 
\begin{eqnarray}
\label{eq:bnpc}
\cos \alpha (T_c) \frac{{\bar\phi} (T_c)}{T_c} \gtrsim 1
\end{eqnarray} 
during the EWPT. If this condition is met, then the EWPT is said to be strongly first-order. We emphasize that, although gauge-independent, Eq.~(\ref{eq:bnpc}) gives a rough BNPC, as there exist a variety of additional theoretical uncertainties that enter the computation of the sphaleron-induced rate for baryon washout (see Ref.~\cite{Patel:2011th} for a detailed discussion).

To evaluate the quantities that characterize the transition, we use the {\tt CosmoTransitions} package~\cite{Wainwright:2011kj} , inputting the finite temperature effective potential in the high temperature limit. Among the primary tasks of the {\tt CosmoTransitions} package is calculating the finite temperature tunneling solution between two different vacua. This tunneling solution characterizes a bubble with an energy $S_3$, the finite temperature, three-dimensional Euclidean action. The ratio of $S_3$ to the nucleation temperature $T_N$ controls the thermal tunneling rate such that nucleation only occurs when $S_3 / T_N \simeq 140$~\cite{Anderson:1991zb}. Although we numerically enforce the electroweak minimum to be the absolute minimum at $T=0$ in our analysis, this does not preclude the existence of other phases at $T=0$ and the possibility that the Universe could become stuck in such a metastable phase. In order to exclude this possibility, we require that a given set of xSM parameters yield $S_3/T_N \simeq 140$ in addition to satisfying Eq.~(\ref{eq:bnpc}).

Before implementing the xSM in {\tt CosmoTransitions}  to calculate all quantities numerically, we include here a short explanation of the finite temperature potential in the high temperature limit for the purposes of relaying physical intuition of our results. At $T=0$, the tree-level potential, \eqn (\ref{TotPot}), can be written in the cylindrical coordinate system as
\begin{eqnarray}
& V_0^{T=0}(\phi, \alpha) = \ds \left( - \mu^2 \cos^2 \alpha + \frac{b_2}{2} \sin^2 \alpha \right) \phi^2 & \nonumber \\
& + \ds \left( \frac{a_1}{2} \cos^2 \alpha + \frac{b_3}{3} \sin^2 \alpha \right) \sin \alpha \phi^3  & \nonumber \\
& + \ds \left( \lambda \cos^4 \alpha + \frac{a_2}{2} \cos^2 \alpha \sin^2 \alpha + \frac{b_4}{4} \sin^4 \alpha \right) \phi^4  .&
\end{eqnarray}
Including the standard one-loop $T\neq 0$ corrections and retaining the leading terms in the high-$T$ expansion, we obtain

%
%
%
%
%
%
%
%
%
\begin{eqnarray}
& \ds V_{eff}^{T \neq 0}(\phi, \alpha, T) = B T^2 \phi + \left( 2 \bar{D} (T^2 - T_0^2) + \frac{b_2}{2} \sin^2 \alpha \right) \phi^2 & \nonumber \\
& + E \phi^3 + \bar{\lambda} \phi^4 &
\end{eqnarray}
with
\begin{eqnarray}
& \ds B = \left( \frac{ a_1 + b_3}{12} \right) \sin \alpha & \nonumber \\
& \ds \bar{D} = D_{SM} \cos^2 \alpha + \frac{1}{48}  \left( a_2 (1+ \sin^2 \alpha ) + 3 b_4 \sin^2 \alpha \right)  & \nonumber  \\
& \ds E = \left( \frac{a_1}{2} \cos^2 \alpha + \frac{b_3}{3} \sin^2 \alpha \right) \sin \alpha & \nonumber \\
& \ds \bar{\lambda} = \lambda \cos^4 \alpha + \frac{a_2}{2} \cos^2 \alpha \sin^2 \alpha + \frac{b_4}{4} \sin^4 \alpha ,& 
\label{FTparams}
\end{eqnarray}
where $D_{SM}$ and $T_0$ correspond to the usual SM values. Note that the $BT^2\phi$ tadpole term is gauge invariant at one-loop order, as it arises from tree level operators. We do not preclude the possibility that gauge-dependence enters at higher order (in contrast to the thermal mass corrections), so we will not retain this term in our analysis\footnote{We also note that it is generally numerically suppressed, given the linear dependence on $\sin\alpha$ and the generally small values of $\alpha$ implied by our study.}.

\begin{figure*}[t!]
\centering
\hspace*{-.3in}\mbox{
\includegraphics[scale=.451]{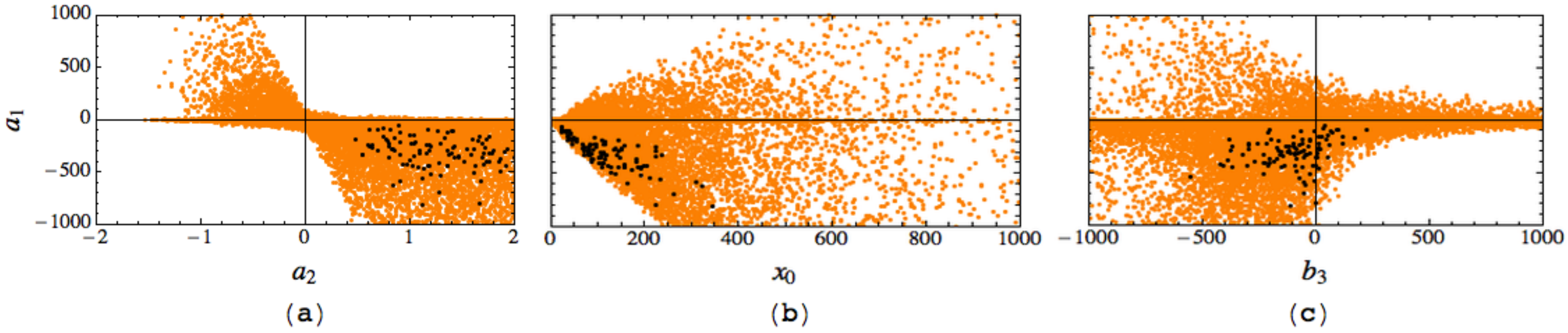}
}
\hspace*{-.2in}\mbox{
\includegraphics[scale=.45]{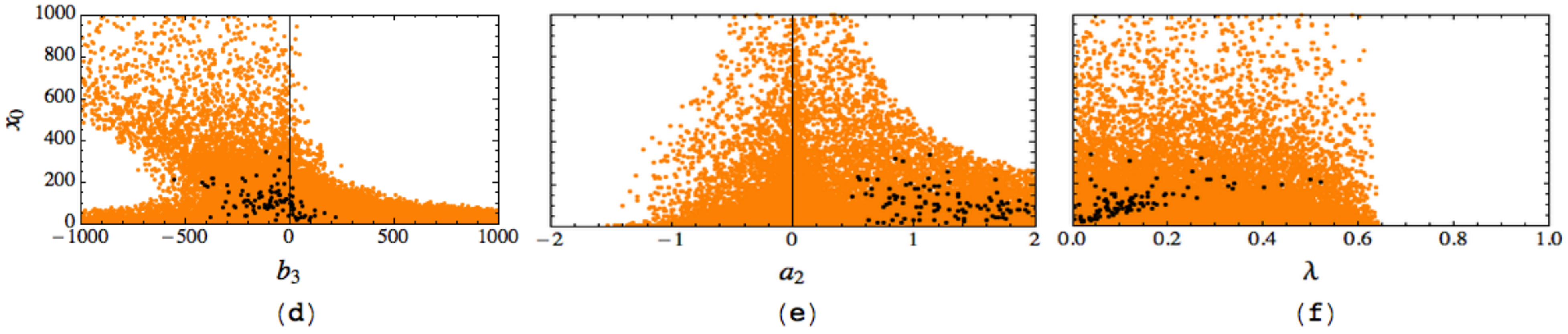}
}
\hspace*{-.2in}\mbox{
\includegraphics[scale=.45]{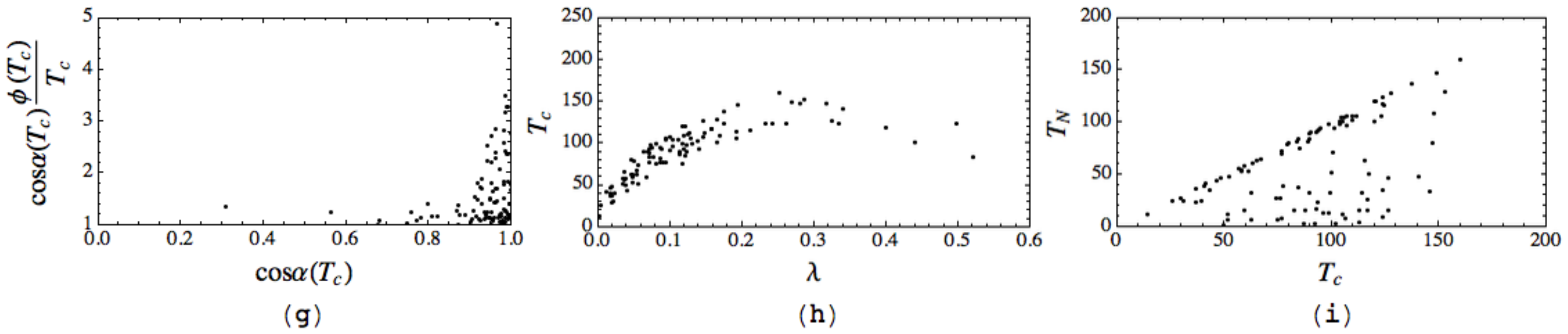}
}
\caption{{\it Scatter plots of the parameter space. Orange (light) points satisfy limits from current LHC measurements of Higgs properties, heavy/light SM-like Higgs searches, and electroweak precision bounds. Black points further satisfy the requirement of a SFOEWPT and exhibit a sufficiently fast thermal tunnelling rate to support bubble nucleation.}}
\label{FIG3}
\end{figure*}
The critical values, $\phi (T_c)$ and $\alpha(T_c)$, are determined by minimizing $V_{eff}^{T \neq 0}(\phi, \alpha, T)$ while $T_c$ is defined by the condition that the broken and unbroken electroweak phases are degenerate, \ie 
\begin{eqnarray}
V_{eff}^{T \neq 0}(\phi, \alpha \neq \pi/2 , T_c) = V_{eff}^{T \neq 0}(\phi, \alpha = \pi/2 , T_c) .
\end{eqnarray}
In terms of the parameters in \eqn (\ref{FTparams}), the condition for a SFOEWPT then becomes
\begin{eqnarray}
- \cos \alpha (T_c) \frac{E (T_c)}{2 T_c \bar{\lambda} (T_c)} \gtrsim 1 .
\label{PTbehaviour}
\end{eqnarray}
Here, we see that a key feature  of the xSM model in relation to the EWPT the generation of a cubic terms in $V_{eff}^{T \neq 0}$ at tree-level via the $\mathbb{Z}_2$-breaking operators. This result agrees with that of~\cite{Profumo:2007wc} modulo the SM loop-generated cubic terms in $V_{eff}^{T \neq 0}$, which we have neglected in our prescription for achieving gauge-independent results\footnote{We again note that we have not included the loop-induced cubic terms whose effect, when evaluated according to the gauge-invariant procedure of Ref.~\cite{Patel:2011th}, will open up additional regions of EWPT-viable parameter space.}.

The interpretation of the effects of the xSM parameters can now be stated as follows: 

\begin{itemize}
\item[$\bullet$] The $\mathbb{Z}_2$-breaking parameters serve to raise the barrier between the electroweak phase and the high temperature phase when the overall combination of $a_1$ and $b_3$ in $E$ is large and negative. 
\item[$\bullet$] The $\mathbb{Z}_2$-conserving quartic parameters in $\bar{\lambda}$ may increase the sphaleron energy by increasing ${\bar v}(T_C)$. They can do this by either all remaining positive and taking on smaller values, or by $a_2$ becoming negative and large enough to reduce the effect of $\lambda$ and $b_4$ while remaining small enough to maintain vacuum stability. Note that vacuum stability dictates that $\lambda$ and $b_4$ both must remain positive. 

We emphasize that the presence of $a_2>0$ can indirectly enable a SFOEWPT by allowing $\lambda$ to take on a smaller magnitude than it would in the SM. In some regions of parameter space, the associated contribution to $m_1^2$ arising from $m_{hs}^2$ allows for a reduction in the contribution from $m_{hh}^2=2 \lambda v_0^2$ while yielding the observed value of the SM-like Higgs mass. The value of $\lambda$ thus required to obtain the observed Higgs-like scalar mass may be smaller than in the SM. Moreover, $\lambda$ enters ${\bar\lambda}$ with a factor of $\cos^4\alpha_c$, so that reducing $\lambda$ can effectively reduce the denominator of Eq.~(\ref{eq:bnpc}) for $|\cos\alpha_c|\sim 1$.

\item[$\bullet$] As our numerical results below indicate, a reduction in the value of $\lambda$ (resulting from $a_2>0$) may also allow for a reduced 
value of $T_c$. 
\end{itemize}

We perform numerical Monte Carlo scans of the xSM parameter space. As our free parameters, we take all cubic and quartic couplings in \eqn (\ref{TotPot}) as well as the $T=0$ singlet vev, providing a full description of the potential. Our scans cover this parameter space within the ranges
\begin{eqnarray}
& a_1/\tev, b_3/\tev \in [-1,1] \qquad x_0/\tev \in [0,1] & \nonumber \\
& b_4, \lambda \in [0,1] \qquad  a_2 \in [- 2 \sqrt{\lambda b_4}, 2] ,&
\end{eqnarray}
where the lower bounds on the quartic couplings represent the vacuum stability bounds presented in section~\ref{sec:Model}. For each point, we require consistency with all bounds from collider searches and electroweak precision observables presented in Fig.~\ref{FIG2}. Moreover, we perform 3 distinct scans, with each scan distinguished by the bound imposed on the mixing angle, $\theta$. The first scan imposes current LHC bounds while the other two scans impose prospective HL-LHC and ILC-3 bounds, respectively. \\

In Fig.~\ref{FIG3}, we present a selection of 2-dimensional slices of the parameter space left after our first  scan, imposing the current LHC bound on the mixing angle. The orange points are compatible with all collider and electroweak precision bounds while the black points further yield a SFOEWPT with the correct thermal tunneling rate for bubble nucleation. In Fig.~\ref{FIG3}(a) we  show the distribution of the Higgs portal parameters, $a_1$ and $a_2$.  At the collider/electroweak precision level, we find that $a_1$ and $a_2$ are strongly anti-correlated, preferring to have opposite signs throughout the space. This preference can be understood from \eqn (\ref{thetabound}), in which the bound on $\sin 2 \theta$ requires that, in the absence of sufficient suppression from the mass splitting, $m_1^2 - m_2^2$, a cancellation between $a_1$ and $2 a_2 x_0$ must occur. Indeed, in the small regions where both $a_1$ and $a_2$ have the same sign, the mass splitting is near its maximum. 

From the standpoint of the SFOEWPT, \eqn (\ref{PTbehaviour}) implies that negative values of $a_1$ are preferred, therefore favoring positive values for $a_2$. This same mechanism is responsible for the correlations seen between $a_1$ and $x_0$ [panel (b)] as well as $x_0$ and $a_2$ [panel (e)]. Moreover, $x_0$ can vary significantly from its value at $T_c$ and, therefore, does not directly enter \eqn (\ref{PTbehaviour}). Instead, it is the SFOEWPT-preferred values for $a_1$, through the bound in \eqn (\ref{thetabound}), that set the scale for the SFOEWPT-preferred values for $x_0$, as evidenced by the approximate linear behaviour in the $a_1$ versus $x_0$ distribution [panel (b)].

\begin{figure*}[t!]
\centering
\hspace*{-.15in}\mbox{
\includegraphics[width=7in,height=1.5in]{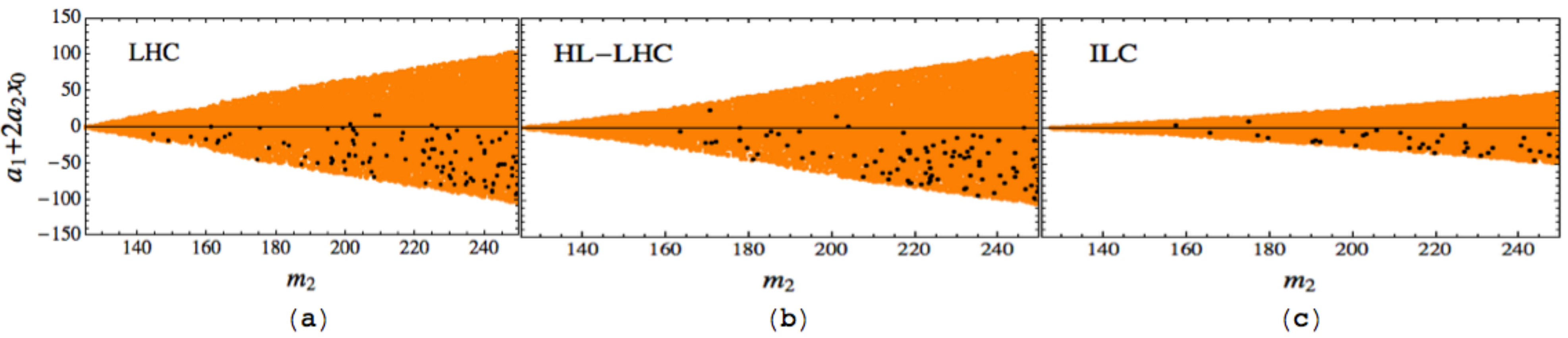}
}
\hspace*{-.05in}\mbox{
\includegraphics[width=7in,height=1.65in]{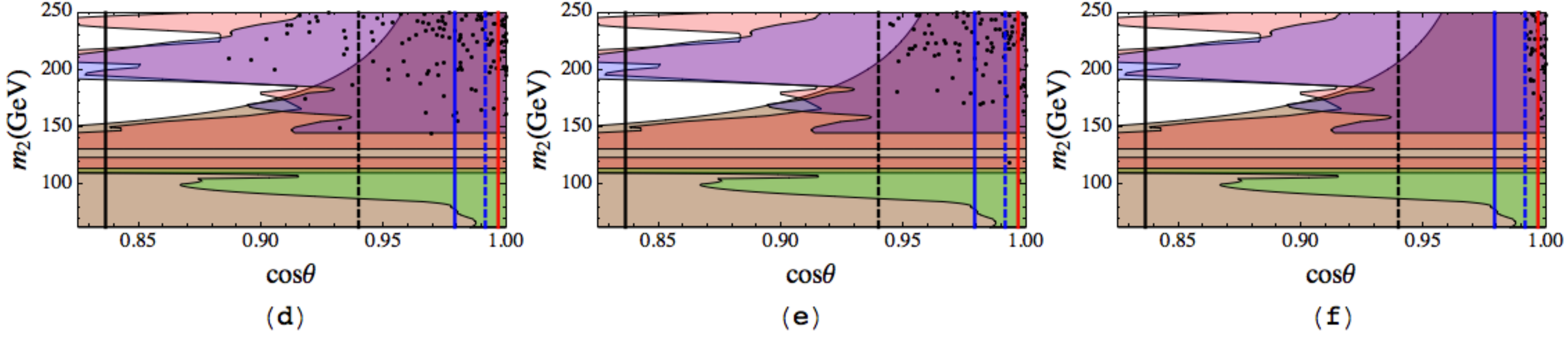}
}
\caption{{\it Top Row: Scatter plots showing the effect of the more stringent bounds on the mixing angle from the HL-LHC and ILC-3 accelerator programs. Bottom Row: Distributions of EWPT-preferred points overlayed with collider and electroweak precision bounds for the current LHC (left), HL-LHC (middle), and ILC-3 (right).}}
\label{FIG4}
\end{figure*}
The correlations seen between $x_0$ and $b_3$ [panel (d)] before imposing the SFOEWPT requirements are less direct, originating primarily from our choice of mass range for $m_2$. Specifically, the addition of the two scalar mass eigenvalues in \eqn (\ref{Meigenvalues}) gives
\begin{eqnarray}
m_1^2 + m_2^2 = m_{hh}^2  + m_{ss}^2\ \ \ .
\end{eqnarray}
Once the relations for $m_{hh}^2$ and $m_{ss}^2$ from \eqn (\ref{mixingM}) are inserted and we make the approximate replacement $a_1 \sim -2 a_2 x_0$, the upper limit, $m_2 < 2 m_1$, takes the form
\begin{eqnarray}
b_3 + 2 b_4 x_0 < \frac{1}{x_0} \left( 5 m_1^2 - 2 \lambda v_0^2 - \frac{a_2}{2} v_0^2 \right) .
\label{b3Bound}
\end{eqnarray}
As $x_0$ becomes large, negative values of $b_3$ are required to remain within the correct mass range. However, as $x_0$ becomes small, $b_3$ is essentially unbounded. Moreover, making the replacement $x_0 \sim \ds \frac{|a_1|}{2 |a_2|}$ in \eqn (\ref{b3Bound}), which accounts for the fact that $a_1$ and $a_2$ are generically of opposite sign, yields
\begin{eqnarray}
b_3 + \frac{b_4 |a_1|}{|a_2|} < \frac{2 |a_2|}{|a_1|} \left( 5 m_1^2 - 2 \lambda v_0^2 - \frac{a_2}{2} v_0^2 \right) .
\end{eqnarray}
This shows that the observed preference for $b_3<0$ in the limit of large $|a_1|$ [panel (c)] also originates from our choice of mass range for $m_2$.
Furthermore, the upper limit observed on $\lambda$ before the EWPT also originates from our choice of mass range. 

From \eqn (\ref{PTbehaviour}), we expect that a SFOEWPT will prefer negative values of $b_3$. However, it is allowed to take on positive values as well due to the fact that its effect on the parameter $E$ is somewhat suppressed by a factor of $\sin^2 \alpha(T_c) / 3$. As shown  in Fig.~\ref{FIG3}(g), a SFOEWPT requires a large $SU_L(2)$ projection (\ie value of $\cos \alpha (T_c)$), implying small $\sin \alpha (T_c)$. This suppression also acts to diminish the effects of $a_2$ and $b_4$ in $\bar{\lambda}$, rendering $\lambda$ the dominant parameter controlling the critical temperature, as shown panel (h). Importantly, the presence of additional contributions to the mass of the SM-like state $h_1$ allows $\lambda$ to take on substantially smaller values than its SM-value $\lambda_\mathrm{SM}\approx 0.13$. Although not all values of $\lambda$ arising from our scan lie below $\lambda_\mathrm{SM}\approx 0.13$, a sizable fraction do fall in this range. Smaller values of $T_C$, in tandem with suitably large values of $E/{\bar\lambda}$, can effectively suppress the sphaleron rate, thereby preventing baryon asymmetry washout.

Finally, in  Fig.~\ref{FIG3}(i), we present the nucleation temperature vs. critical temperature. Points along the diagonal represent a difference of $~T_c-T_N {\sim}$5 \gev~while points below the diagonal represent varying levels of supercooling. All points lie above $T_N {\sim}$5 GeV, and so the model is safe from BBN constraints. However, such supercooling effects result in a further enhancement of EWPT strength by a factor of $T_c / T_N$ that is quite significant for some parameter space points.


\subsection{Phenomenological implications}

We now consider the phenomenological implications of a SFOEWPT, concentrating first on the 
effect on the parameter space by subjecting the mixing angle to the more stringent, prospective HL-LHC and ILC bounds\footnote{Due to the proximity of the ILC-3 and TLEP mixing angle bounds, we do not explicitly consider the effect of TLEP bounds here.}. In doing so, we assume the future measurements will be consistent with a pure SM-like Higgs boson. We defer an analysis of discovery potential, corresponding to a value of $\cos\theta$ differing from unity by more than five standard deviations, to future work. By concentrating on the prospective exclusion, we nevertheless illustrate the reach of future precision Higgs studies in probing the possibility of a SFOEWPT in the xSM.

Under this set-up, we find that for all the 2-dimensional slices of the parameter space shown in Fig.~\ref{FIG3}, the effect of performing a scan with more stringent constraints on the mixing angle is to simply reduce the density of points. As the shape and nature of the correlations do not significantly change, we do not present the full parameter space again. However, in the top row of Fig.~\ref{FIG4}, we show that the impact of prospective  HL-LHC and ILC-3 bounds on the parameter space [panels (b) and (c), respectively]  is to extend the efficiency of the cancellation between $a_1$ and $2 a_2 x_0$ to higher values of $m_2$, corresponding to narrower \lq\lq cones" of orange points. This enforces successively higher levels of tuning throughout the parameter space, making it more difficult to achieve a strong SFOEWPT and therefore universally reducing the point density. \\

In the bottom row of Fig.~\ref{FIG4}, we present the distributions of points yielding a SFOEWPT and sufficiently large thermal tunneling rates in the $m_2$ vs. $\cos \theta$ space for the three separate numerical scans, overlaying the resulting distributions on top of the constraints derived in section~\ref{sec:ColliderPhenoEW}. We observe a tendency for points to prefer larger masses and smaller mixing angles. Specifically, in the case of the current LHC [panel (d)], more than half the black points lie in the region $m_2 > 200 \gev$ and $\cos \theta > 0.97$. This tendency is far more than necessary from any constraint from current collider bounds or electroweak precision observables, so we infer that it is EWPT-induced. 


In order to understand this preference for values of $m_2$ near the upper end of the mass range considered here, 
we note 
a SFOEWPT prefers large negative values of $a_1$, potentially making a perfect cancellation between $a_1$ and $2 x_0 a_2$  in \eqn (\ref{thetabound}) difficult. In the case of an imperfect cancellation, large mass splitting between the two scalar eigenstates is required to compensate and remain within this bound. Moreover, both large mass splitting and efficient cancellations drive the parameter space towards small mixing angles.  This phase-transition induced tendency of the parameter space can be viewed as cosmologically driven motivation for direct searches for new low mass ($\lesssim 2 m_h$) scalar states as well as high precision measurements of Higgs signal strengths at the HL-LHC and ILC.
The ILC, TLEP, and CEPC -- with their exceedingly stringent projected sensitivity to the mixing angles -- hold considerable promise for observing non-zero mixing associated with the SFOEWPT-viable parameter space.  \\

It is also interesting to consider the implications of future measurements of the Higgs-like boson trilinear self-coupling, as suggested by the early analysis of Ref.~\cite{Noble:2007kk}. For center of mass energies below the di-Higgs production threshold, an indirect determination can be obtained through measurements of the Higgs associated production cross section in $e^+e^-$ annihilation\cite{McCullough:2013rea}. Assuming a 0.4\% determination of this cross section at TLEP 240 or the CEPC, one may infer a value of the self-coupling with $\sim$ 30\% precision. With an upgrade to $\sqrt{s}=$500 \gev~and 1 ab$^{-1}$ of luminosity, a 50\% direct determination may be possible at TLEP~\cite{Gomez-Ceballos:2013zzn} using di-Higgs production. Projections for the HL-LHC (again using di-Higgs production) range from 50\% for the $b{\bar b}\gamma\gamma$ channel~\cite{Yao:2013ika} to 30\%, assuming other channels such as $b{\bar b}W^+W^-$ and $b{\bar g}\tau^+\tau^-$ can be measured with similar precision (see \cite{Gomez-Ceballos:2013zzn} and references therein). At the ILC, a combination using the $e^- e^+ \to Zhh$ and $e^- e^+ \to \nu \bar{\nu} hh$ channels may allow for a 13\% determination~\cite{Asner:2013psa}. The most promising scenarios are for a 100 TeV $pp$ collider, for which projections fall in the 5-8\% range~\cite{Gomez-Ceballos:2013zzn,Yao:2013ika}. 

In the xSM, the self coupling of the SM-like Higgs boson is given by the quantity
\begin{eqnarray}
\label{eq:self}
g_{111} = \lambda v_0\cos^3\theta +\frac{1}{4}(a_1+2 a_2 x_0)\cos^2\theta\sin\theta\\
\nonumber
+\frac{1}{2}a_2v_0\cos\theta\sin^2\theta+\frac{b_3}{3}\sin^3\theta+b_4 x_0\sin^3\theta\ \ \ .
\end{eqnarray}
Given the small values of $\theta$ preferred by the present phenomenological constraints, the tendency toward negative values of $(a_1+2 a_2 x_0)$ as indicated by the top row of Fig.~\ref{FIG4}, and the concentration of points for $\lambda < \lambda_{SM}$ as given in Fig.~\ref{FIG3}(h), we expect $g_{111}$ to accommodate values significantly below its SM value $\lambda_{SM}v_0\approx 33$ GeV. 

\begin{figure}[t!]
\centering
\includegraphics[scale=.675]{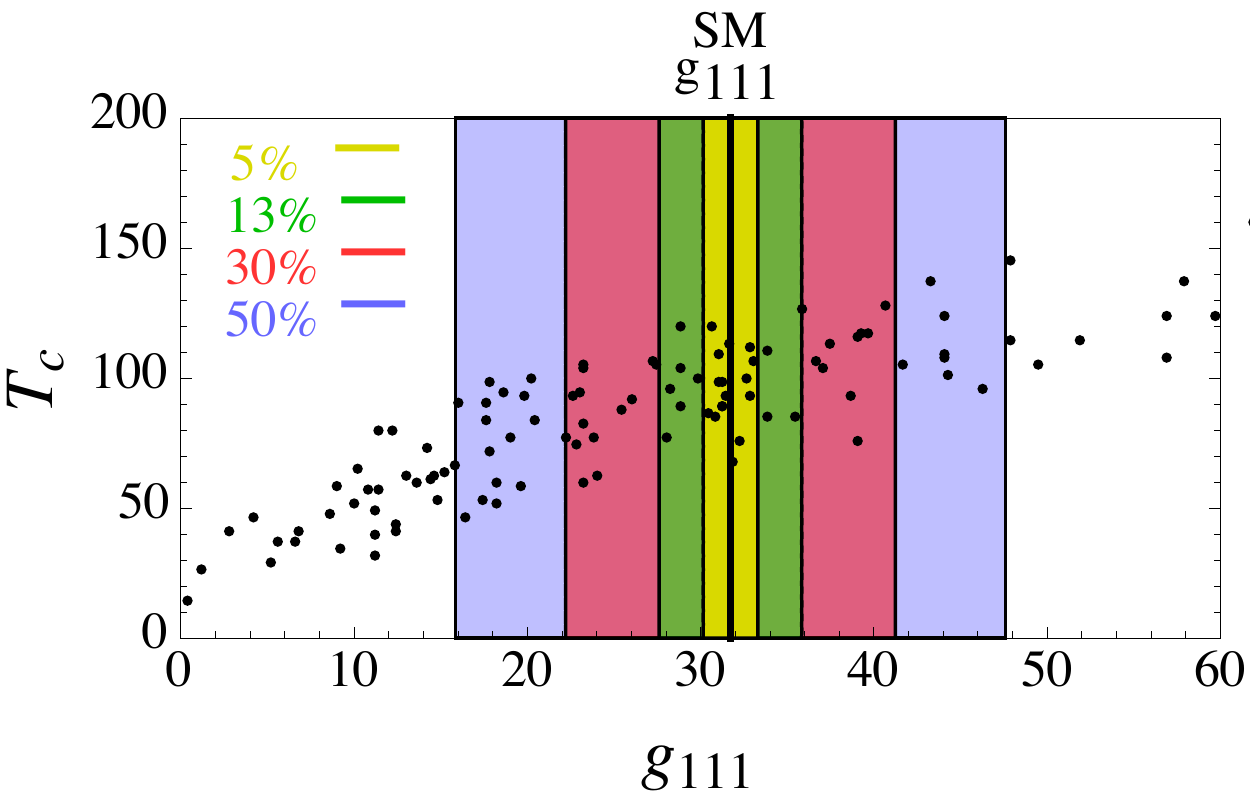}
\caption{\textit{Correlation between the SM-like scalar $(h_1)$ self-coupling $g_{111}$ and the critical temperature for SFOEWPT-viable parameter space points. Blue, red, green, and yellow bands represent, respectively, a $\pm$50\%, $\pm$30\%, $\pm$13\%, and $\pm$5\% variation in $g_{111}$ about its SM value.}}
\label{fig:g111}
\end{figure}
In Fig.~\ref{fig:g111}, we show the correlation between $g_{111}$ and the critical temperature. As expected, we observe that (a) this correlation largely parallels the correlation between $\lambda$ and $T_c$; (b) a substantial fraction of the SFOEWPT parameter choices allow for a reduction in $g_{111}$ from its SM value; and (c) decreasing $g_{111}$ implies decreasing $T_c$. An increase in $g_{111}$ over the SM value by as much as a factor of two or more may also be possible. Thus, a precise determination of $g_{111}$ would provide a powerful probe of the SFOEWPT-viable parameter space. To illustrate this potential, we show in Fig.~\ref{fig:g111} bands corresponding to $\pm 50\%$, $\pm 30\%$, $\pm 13\%$, and $\pm 5\%$ variations in $g_{111}$ about its SM value corresponding roughly to the prospective future collider sensitivities summarized above. We see that there exists a non-negligible fraction of the SFOEWPT-viable points that would lead to significant and observable deviations from the SM expectations for $g_{111}$, particularly with the precision expected for the full ILC data set and the VHE-LHC or SPPC. Conversely, agreement with the SM value could yield stringent constraints on the possibility of a SFOEWPT in this scenario.


\section{Conclusions}
\label{sec:Conclusions}


Uncovering the dynamics of EWSB in the early universe and its possible connection with the origin of the baryon asymmetry remains a key task in particle physics. While the SM scalar sector does not allow for out-of-equilibrium dynamics needed for baryogenesis, simple extensions of the scalar sector can accommodate a SFOEWPT as required by electroweak baryogenesis scenarios. In this paper, we have revisited the implications for the collider phenomenology and the EWPT of the simplest extension of the SM scalar sector containing one additional real gauge singlet scalar field, or xSM. This model exemplifies the phase transition dynamics  of more extensive SM-extensions incorporating gauge singlet scalars, \eg variants of the minimal supersymmetric SM that include a singlet superfield.  Focusing on the kinematic regime in which no new scalar decay modes arise, we have updated the constraints on the parameters of the xSM in light of the discovery of a Higgs-like scalar at the LHC and present determinations of its signal strengths. We have then shown how there exist considerable regions of SFOEWPT-viable parameter space that one could probe with future precision Higgs studies at the HL-LHC, ILC, TLEP, CEPC, VHE-LHC and/or SPPC as well as with searches for singlet-like scalars in the low mass region, $<2 m_h$. 

Should future experiments find evidence for non-zero Higgs-singlet mixing, a substantial deviation of the Higgs trilinear self-coupling from its SM value, and the existence of a second singlet-like scalar having SM-Higgs branching ratios, our analysis would then allow one to narrow down the regions of xSM parameter space consistent with a SFOEWPT. A quantitatively robust assessment of the viability of such a transition and a determination of its characteristics would then require a Monte Carlo study, given the limitations of perturbation theory in this context (for a discussion of these limitations, see \eg Ref.~\cite{Patel:2011th}). The outcome of such a program would constitute a significant step toward explaining the abundance of visible matter in the universe.

\acknowledgements We thank H. Patel for collaboration during the early stages of this work and T. Liu, D. Stolarski,  J. Shu for helpful discussions. SP, MJRM and PW also thank the Kavli Institute for Theoretical Physics where a portion of this work was completed. This work was supported in part by U.S. Department of Energy contracts DE-SC0011095 (MJRM and PW), DE-FG02-04ER41268 (SP), and by the National Science Foundation under Grant No. NSF PHY11-25915 (SP, MJRM, and PW).

\end{document}